\documentclass{sf2a-conf2017}
\usepackage{graphicx}
\usepackage[]{hyperref}
\usepackage[]{natbib}  
\usepackage{epstopdf}
\usepackage{sidecap}
\usepackage{wrapfig,lipsum,booktabs}

\def\BibTeX{{\rm B\kern-.05em{\sc i\kern-.025em b}\kern-.08em
    T\kern-.1667em\lower.7ex\hbox{E}\kern-.125emX}}
\bibpunct{(}{)}{;}{a}{}{,}  

\begin{document}

\TitreGlobal{SF2A 2017}


\title{A QUESTION OF MASS : ACCOUNTING FOR ALL THE DUST IN THE CRAB NEBULA WITH THE DEEPEST FAR INFRARED MAPS}

\runningtitle{Dust in the Crab Nebula}

\author{J. Matar}\address{Department of Physics \& Astronomy, Notre Dame University, Lebanon; cnehme@ndu.edu.lb}

\author{C. Nehm\'e$^{1,}$}\address{Laboratoire AIM, CEA/IRFU/Dep. d'Astrophysique, CNRS, Universit\'e Paris Diderot, 91191 Gif-sur-Yvette, France}

\author{M. Sauvage$^2$}

\setcounter{page}{237}

\maketitle

\begin{abstract}
Supernovae represent significant sources of dust in the interstellar medium. In this work, deep far-infrared (FIR) observations of the Crab Nebula are studied to provide a new and reliable constraint on the amount of dust present in this supernova remnant. Deep exposures between 70 and 500 $\mu$m taken by PACS and SPIRE instruments on-board the Herschel Space Telescope, compiling all observations of the nebula including PACS observing mode calibration, are refined using advanced processing techniques, thus providing the most accurate data ever generated by Herschel on the object. We carefully find the intrinsic flux of each image by masking the source and creating a 2D polynomial fit to deduce the background emission. After subtracting the estimated non-thermal synchrotron component, two modified blackbodies were found to best fit the remaining infrared continuum, the cold component with T$_c$ = 8.3 $\pm$ 3.0 K and M$_d$ = 0.27 $\pm$ 0.05 M$_{\odot}$ and the warmer component with T$_w$ = 27.2 $\pm$ 1.3 K and M$_d$ = (1.3 $\pm$ 0.4) $\times$10$^{-3}$ M$_{\odot}$.
\end{abstract}


\section{Introduction}
It is well established that dust is efficiently formed in regions
around asymptotic giant branch (AGB) stars. Thus AGB stars are classically considered as the primary source of dust grains in galaxies. Nevertheless, they cannot inject enough dust in the interstellar medium to compensate for the known destruction rates. Supernova (SN) explosions in the interstellar medium (ISM) trigger shock waves that are considered the dominant mechanism of dust destruction. Yet there is increasing  evidence  for the formation of non-negligible quantities of dust grains in the
ejecta of SNe \citep{Bocchio}. The Spitzer and Herschel space telescopes have thus observed supernovae remnants (SNRs), currently known to be major sources of dust. Such is the case of the Crab Nebula, a young, bright and well resolved SN at a distance of 2 kpc \citep{Trimble}, where large amounts of dust have been reported. Unlike other remnants, there is almost no interstellar material in front of or behind the Crab Nebula, so the possibility that the observed dust is simply the local ISM swept up by the expanding shockwave is disfavoured. Therefore, it represents an ideal case for deriving the total dust mass produced by a SNR. Previous studies using Herschel deduced masses ranging between 0.12-0.25 $M_{\odot}$ \citep{gomez2012} and 0.019-0.13 $M_{\odot}$ \citep{tea2013}. These results consistently indicate that dust is formed in the ejecta, but their variability shows that much work is still needed to get a reliable estimate of the dust yields by SN. We also note that the Crab Nebula is ionized by non-thermal radiation \citep{MacAlpine} thus measuring accurately the synchrotron emission is important to correctly derive the dust thermal emission.

\section{Observations}
\begin{figure}[h]
\centering
\includegraphics[width=0.92\textwidth,clip]{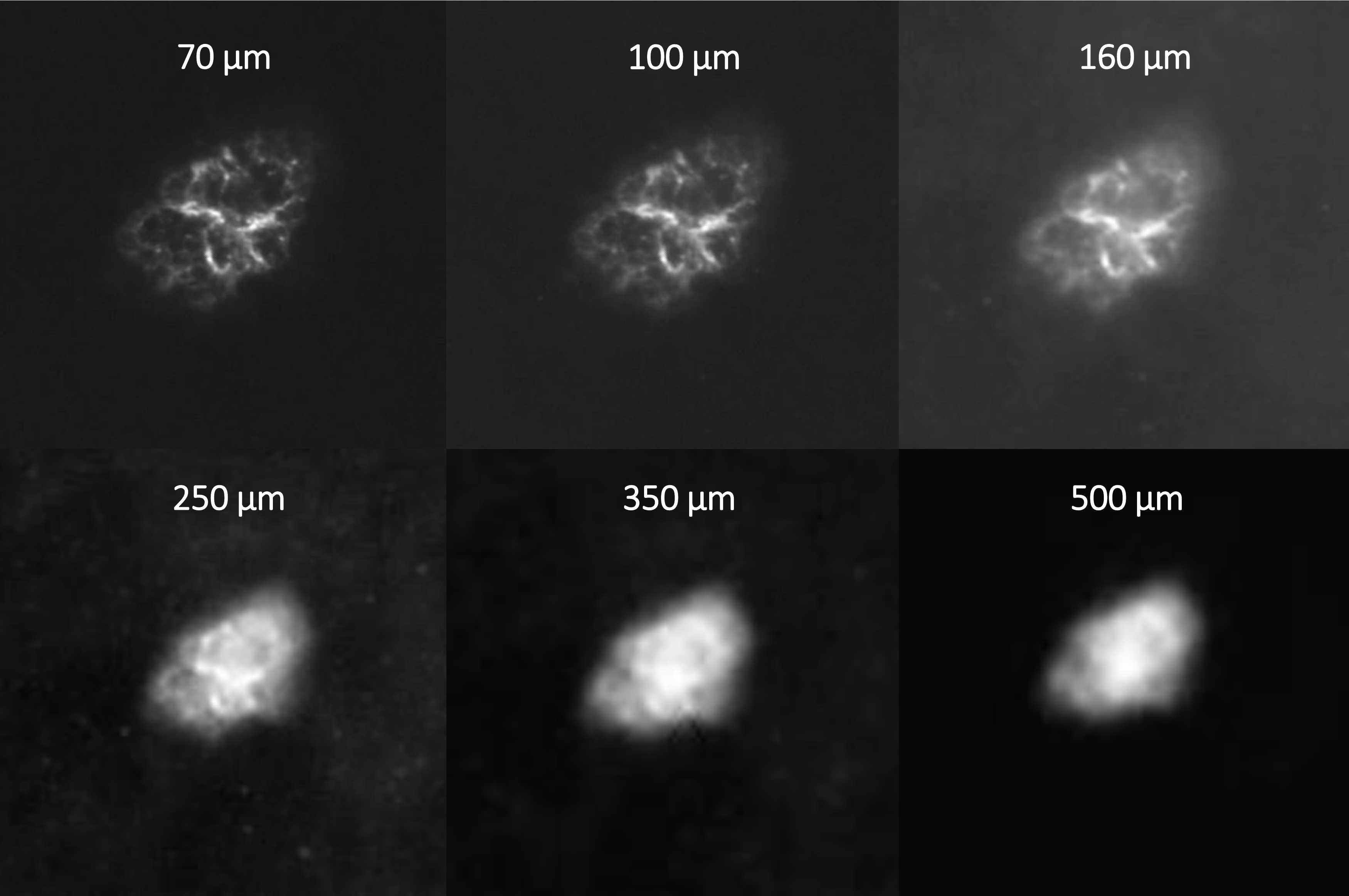}
\caption{\textit{Top:} Herschel PACS 70, 100 and 160 $\mu$m images. \textit{Bottom:} Herschel SPIRE 250, 350 and 500 $\mu$m images.}
\label{Figure: fig1}
\end{figure}

The Crab Nebula was observed using Herschel \citep{herschel2010} between September 2009 and September 2010. The PACS \citep{PACS2010} and SPIRE \citep{SPIRE2010} instruments performed photometry (Figure \ref{Figure: fig1}) at 70, 100, 160, 250, 350 and 500 $\mu$m (see Table \ref{Table: Observations}). The PACS  data were first reduced using HIPE, the Herschel Interactive Processing Environment \citep{Ott2010}. Then the ``unimap" software \citep{Piazzo2015}, a generalized least-squares map maker for Herschel data, was used to get rid of low frequency noise in the maps. All PACS observations were combined in one map per photometer band.
For SPIRE, the standard pipeline HIPE version 14.0.0 was used \citep{SPIRE2010}.
Note that in this study, we have included for the first time data that was obtained on the Crab Nebula during the testing and qualification of PACS observing modes, thus nearly doubling the depth of the resulting maps.

\begin{table}[h]
\centering
\caption{Photometric observations of the Crab Nebula using Herschel Space Telescope.}
\begin{tabular}{c c c c c}
\hline
\hline
\textbf{Obs. ID} & Obs. Date & Obs. Duration (s) & \textbf{Instrument} & Filter set ($\lambda$ ($\mu$m) \\
\hline
\hline
1342183905 & 2009-09-15 & 2221 & PACS & (70, 160) \\
1342183906 & 2009-09-15 & 2221 & PACS & (70, 160) \\
1342183907 & 2009-09-15 & 2221 & PACS & (70, 160) \\
1342183908 & 2009-09-15 & 2221 & PACS & (70, 160) \\
1342183909 & 2009-09-15 & 2221 & PACS & (100, 160) \\
1342183910 & 2009-09-15 & 2221 & PACS & (100, 160) \\
1342183911 & 2009-09-15 & 2221 & PACS & (100, 160) \\
1342183912 & 2009-09-15 & 2221 & PACS & (100, 160) \\
1342191181 & 2010-02-25 & 4555 & SPIRE & (250, 350, 500) \\
1342204441 & 2010-09-13 & 1671 & PACS & (70, 160) \\
1342204442 & 2010-09-13 & 1671 & PACS & (100, 160) \\
1342204443 & 2010-09-13 & 1671 & PACS & (70, 160) \\
1342204444 & 2010-09-13 & 1671 & PACS & (100, 160) \\
\hline
\end{tabular}
\label{Table: Observations}
\end{table}

\section{Analysis}
\subsection{Aperture Photometry and background subtraction}
To build the overall SED of the Crab Nebula, we performed aperture photometry on the maps. We defined a single aperture that contains the full extent of the Crab Nebula at all observed wavelengths. Since the apparent size of an object increases with the wavelength when observed with the same telescope, we define the aperture on the longest wavelength image, the 500 $\mu$m map. The same circular aperture was used for the six images.
In the infrared, there is always more emission than simply the objects we are interested in. Any pixel inside the aperture will see flux from the object and flux from the background. Therefore, we need to estimate the background level and subtract it. For that purpose, we designed a 2D Polynomial Fit Model to estimate the background from the emission outside the circular aperture.
\subsection{Synchrotron Emission}
The Crab Pulsar B0531+21 is the strongest source of synchrotron radiation in the Galaxy. We estimated the non-thermal synchrotron component by applying a least-squares fit to the Herschel (our data), Spitzer, WISE and Planck data from \citep{gomez2012}. This produces a power law with a spectral index $\alpha$ = 0.413 $\pm$ 0.085.

\section{Results}
\subsection{Flux Determination}
In table \ref{Table: Fluxes}, we list the fluxes extracted from our maps using the analysis method presented above. 
\begin{table}[h]
\centering
\caption{The background and synchrotron contribution to the fluxes.}
\begin{tabular}{c c c c c c}
\hline
\hline
\textbf{$\lambda$ ($\mu$m)} & \textbf{$S_{Total}$} (Jy) & \textbf{$S_{Background}$} (Jy) & \textbf{$S_{Synchrotron}$} (Jy) & \textbf{$S_{\nu}$} (Jy) & \textbf{$\Delta\,S_{\nu}$} (Jy)\\
\hline
70 $\mu$m & 220.696 & 2.523 & 45.719 & 172.454 & 0.186\\ 
100 $\mu$m & 225.227 & 5.189 & 52.992 & 167.047 & 0.178\\ 
160 $\mu$m & 153.456 & 28.014 & 64.369 & 61.073 & 0.349\\ 
250 $\mu$m & 214.253 & 104.637 & 77.427 & 32.189 & 0.217\\ 
350 $\mu$m & 170.179 & 66.038 & 88.995 & 15.147 & 0.214\\ 
500 $\mu$m & 133.745 & 29.172 & 103.15 & 1.423 & 0.129\\
\hline
\end{tabular}
\vspace{0.5cm}
\textbf{\\Notes.} $S_{\nu}$ is the total dust emission and $\Delta\,S_{\nu}$ is the photometric error due to background subtraction.
\label{Table: Fluxes}
\end{table}

\subsection{Spectral Energy Distribution}
\begin{SCfigure}
\includegraphics[scale=0.36]{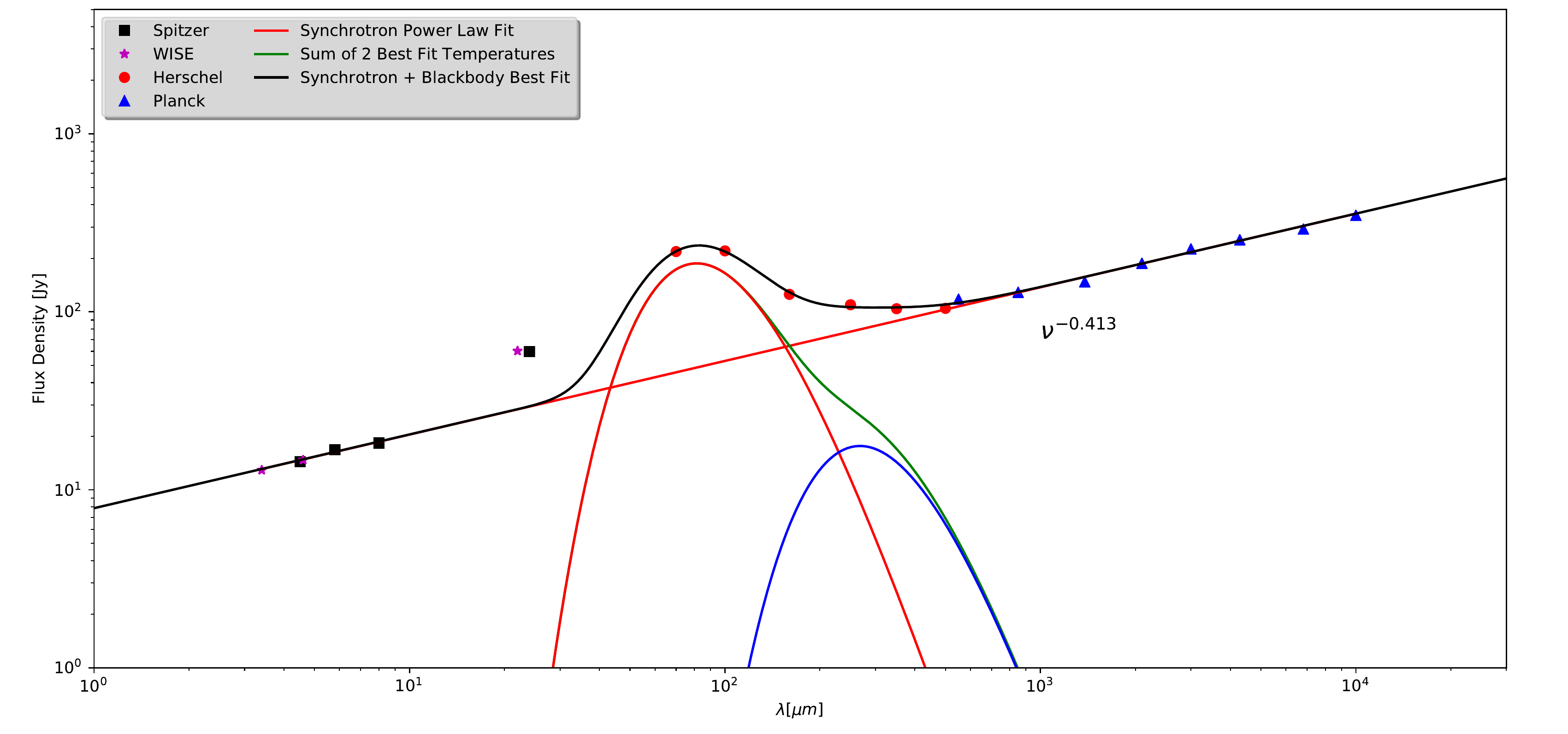}
\caption{SED of the Crab Nebula from the IR-radio including Herschel (red points) and Spitzer (black), WISE (purple) photometry
and Planck fluxes (Planck Collaboration 2011, blue points). The black line is the total flux obtained from summing the synchrotron and the MBB.}
\label{Figure: fig2}
\end{SCfigure}

If all dust grains were to share the same size and composition \citep{Bianchi}, the emission of the Crab would be that of a modified blackbody (MBB), i.e. a blackbody multiplied by the dust absorption cross section. We use here this simplified approach to model the excess thermal emission observed in the FIR (Figure \ref{Figure: fig2}), by the sum of two modified blackbodies identified with a warm and a cold component. The figure shows that the Spitzer and WISE fluxes, consistent with one another, are not reproduced by our fit. This is expected as our model does not represent very small grains able to reach, transiently, very high temperature. We defer this to future work where we will implement a more complex representation of the dust. For all the other data points, the agreement is excellent, and the best-fitting temperatures are shown in Table~\ref{Table:Results}.

\subsection{Total Dust Mass}
From \citep{Laor1993} and \citep{Weingartner2001}, the dust mass for each component is:

\begin{equation}
M_d = \frac{S_{\nu} D^2}{\kappa_{\nu} B(\nu,T)}
\label{Md}
\end{equation}

\noindent where, $S_{\nu}$ is the flux density, D is the distance (Earth-Crab Nebula), $B(\nu, T)$ is the Planck function and $\kappa_{\nu}$ is the dust mass absorption coefficient.

\begin{table}[h]
\centering
\caption{Result from the two-component modified black-body fit}
\begin{tabular}{l c c}
\hline
\hline
Component & Temperature (K) & Mass (M$_{\odot}$) \\
\hline
\hline
Warm & $27.3\pm1.3$ & ($1.3\pm0.4)\times10^{-3}$\\
Cold & $8.2\pm3.0$ & $0.27\pm0.05$\\
\hline
\end{tabular}
\label{Table:Results}
\end{table}

\section{Conclusion}
In this work we present maps of the Crab Nebula with double the depth of previous PACS maps. We rederived the mass of dust using a modified black body with two temperature components fitted to the global SED. We report masses of M$_d(cold)$ = 0.27 $\pm$ 0.05 M$_{\odot}$, M$_d(warm)$ = 1.3 $\pm$ 0.4 $\times$10$^{-3}$ M$_{\odot}$. Our results represent additional support to  theoretical models predicting a total dust mass between 0.2 and 0.5 M$_{\odot}$ (\cite{Woosley1995} ; \cite{Limongi2003}). \cite{Todini2001} also predict between 0.1 and 0.3 M$_{\odot}$ of dust should form in the ejecta from the (Type-IIP) explosions of progenitor stars with initial mass $<$ 15 M$_{\odot}$ such as the Crab progenitor. Nevertheless, this work is a first step of a long term project that we started on dust in the Crab nebula, to better account for the actual complexity of the source (i.e: filaments) by using the spatial information as well as the spectral information.

\begin{acknowledgements}
\textbf{Acknowledgement}\\
Dott. Elias Kammoun ... Thank you!
\end{acknowledgements}

\bibliographystyle{aa}  
\bibliography{sf2a-template} 

\end{document}